\documentclass[prb,showpacs,twocolumn,superscriptaddress,aps,a4paper]{revtex4}

\usepackage{dcolumn}
\usepackage{amsmath}
\usepackage{graphicx}
\usepackage{latexsym}
\usepackage{amsfonts}
\usepackage{amssymb}

\def\a{\alpha}
\def\b{\beta}
\def\d{\delta}
\def\F{\Phi}
\def\g{\gamma}
\def\G{\Gamma}

\def\q{\psi}
\def\Q{\Psi}
\def\r{\rho}

\def\S{\Sigma}
\def\t{\tau}

\def\bra{\langle}
\def\ket{\rangle}

\def\ra{\rightarrow}
\def\inf{\infty}
\def\de{\partial}

\newcommand{\be}{\begin{equation}}
\newcommand{\ee}{\end{equation}}
\newcommand{\beq}{\begin{eqnarray}}
\newcommand{\eeq}{\end{eqnarray}}

\begin{document}

\title{Wick Theorem for General Initial States}

\author{R. van Leeuwen}
\affiliation{Department of Physics, Nanoscience Center, 
FIN 40014, University of Jyv\"askyl\"a, Jyv\"askyl\"a, Finland}
\affiliation{European Theoretical Spectroscopy Facility (ETSF)}

\author{G. Stefanucci}
\affiliation{Dipartimento di Fisica, Universit\`a di Roma Tor
Vergata, Via della Ricerca Scientifica 1, 00133 Rome, Italy}
\affiliation{Istituto Nazionale
di Fisica Nucleare, Laboratori Nazionali di Frascati, Via E. Fermi 40, 00044 Frascati, Italy}
\affiliation{European Theoretical Spectroscopy Facility (ETSF)}

\begin{abstract}
We present a compact and simplified proof of a generalized Wick theorem to calculate the Green's 
function of bosonic and fermionic systems in an arbitrary  
initial state. It is shown that the decomposition of the 
non-interacting $n$-particle Green's function is equivalent to 
solving  a boundary problem for the Martin-Schwinger hierarchy; 
for non-correlated initial states a one-line proof of the standard 
Wick theorem is given. Our result leads to new self-energy 
diagrams and an elegant relation with those of the imaginary-time 
formalism is derived. The theorem is easy to use and can be combined 
with any ground-state numerical technique to calculate time-dependent properties.
\end{abstract}


\maketitle

\section{Introduction}

The theory of Green's functions is probably the most powerful and versatile 
formalism in physics. Due to its generality it has found widespread
applications in any branch of physics dealing with many-particle systems
such as in nuclear physics, condensed matter physics and atomic and molecular physics.
The key of its success lies in  the possibility of 
expanding the dressed (interacting) $n$-particle Green's 
function $G_{n}$ in 
terms of the bare (non-interacting) Green's functions $g_{m}$, 
$m\geq n$, 
which are then reduced to an (anti)symmetrized product of $g_{1}$ 
by means of the Wick theorem.\cite{w.1950}
In his memorable talk 
%at the Pocono Manor Inn
in 1948 Feynman showed how to represent the cumbersome Wick expansion 
in terms of physical insightful diagrams, and since then the Feynman diagrams became an 
invaluable tool  in many areas of physics.

The standard Green's function formalism (GFF) is, however, applicable  to 
non-degenerate systems initially in their ground state and cannot be 
straightforwardly applied to systems 
with more general initial states or to systems with degenerate ground states. 
This leaves out modern fields of research like, e.g., non-equilibrium phase 
transitions,\cite{o.2004}
relaxation dynamics of ultracold gases,\cite{cr.2010}
response of nanoscale systems in non-equilibrium 
steady-states,\cite{mssls.2008} optimal control theory,\cite{bcr.2010} etc. 
The correct description of initial correlations is obviously crucial 
for the short-time dynamics of general quantum systems such as in
transient dynamics in quantum transport\cite{mssls.2008,myoh.2009} or in the study of
atoms and molecules in external laser fields.\cite{dahlen.2007,balzer.2010}
In the case of finite systems it is clear that initial correlations also affect the long-time 
dynamics due to the presence of discrete quantum numbers. However, also in extended systems
initial correlations can have infinite memory when the underlying 
Hamiltonian is integrable. This has, for instance,
recently been demonstrated for interaction quenches
\cite{cazalilla,perfetto1,cazalilla2,uhrig,zhou,dhz.2010,perfstef.2010}
and quantum transport \cite{perfstef.2011} in Luttinger liquids. 
Another case in which the standard GFF is problematic, even for 
equilibrium properties, is that
of systems with degenerate ground states such as in open shell atoms or molecules.
These systems are ubiquitous in quantum chemistry but the standard Wick theorem can only deal
with the highest spin- or angular momentum component of a multiplet.\cite{cederbaumschirmer.1974}
There is therefore a clear need to go beyond the standard GFF.

In 1975 Hall \cite{h.1975} 
%as well as later work \cite{Henning90,Fauser95} 
used the textbook expansion of time-ordered products of operators into 
normal-ordered products and contractions to extend the Wick theorem to arbitrary 
initial states. Although the general structure is outlined there, the 
calculations of the various prefactors is very laborious as it 
requires the explicit determination of the sign of all possible 
permutations of the $2m$ indices of the $g_{m}$'s.
Further progress was made in the classic review of Danielewicz\cite{d.1984} in which it was shown
that one can deal with arbitrary initial states by
introduction of an extended Keldysh contour with an additional 
imaginary track on which the
initial density matrix is represented as the exponential of an, in general, $n$-body operator. 
This was worked out further in detail by Wagner \cite{w.1991} who derived 
the Feynman rules for the Green's function involving
diagrams with arbitrary $n$-body correlators. Later the approach 
was used by Mozorov and R\"{o}pke \cite{mr.1999}
to derive quantum kinetic equations with arbitrary initial correlations.
Furthermore Bonitz and co-workers have presented alternative derivations of the equations
of motion of the non-equilibrium Green's functions with arbitrary initial correlations using
functional derivative techniques and applied them to study the decay of initial correlations
in an electron gas. \cite{skb.1999} Recent work has further explored how 
an initial density matrix on the extended Keldysh contour can be related
to diagrammatic expansions on the standard Keldysh contour.\cite{garny.2009}

The aim of the present work is to extend these developments in three different ways.
First of all, we recognize that the generalized Wick theorem is simply
the solution of the coupled system of differential equations for the 
$g_{m}$'s, the so called 
Martin-Schwinger hierarchy (MSH),\cite{ms.1959} with proper boundary conditions. 
This completely avoids the introduction of normal-ordered products or contractions
and therefore greatly simplifies the mathematics.
In particular the standard Wick theorem follows as a one-line proof.
Secondly, our reformulation based on an initial value problem allows us to 
prove that the generalized Wick expansion has a form identical to that of a 
Laplace expansion for permanents/determinants  (for bosons/fermions). Consequently, the calculation 
of the various prefactors is both explict and greatly simplified.
We provide a systematic way to express the $g_{m}$ in terms of 
$g_{1}$ and of the initial $k$-particle density matrices, $k\leq m$. 
The latter encode all the necessary information on the initial state 
and are completely determined by it.
When the initial state is a permanent/determinant of single particle 
states the $g_{m}$ is expressed solely in terms of the $g_{1}$ 
(standard Wick theorem).
Novel terms, instead, appear  for initially correlated or entangled 
states. Such terms have a simple diagrammatic representation and 
are easier to evaluate if compared to  standard diagrams.
Thirdly, we discuss the relation between GFF based on the generalized 
Wick theorem and GFF as formulated on the extended Keldysh contour.
These two approaches are complementary: the former requires the knowledge of the initial 
state while the latter requires the knowledge of the Hamiltonian
with such initial state as the ground state. Clearly the convenience of using
one approach or the other depends on the information at hand. 
We present a straightforward proof of the existence of a  
Dyson equation for general initial states and derive an 
exact mathematical relation between its self-energy and the 
self-energy of GFF on the extended Keldysh contour.

\section{The standard Green's function formalism}

Let us briefly review under which circumstances the Wick theorem can be 
used to calculate the $n$-particle Green's function $G_{n}$.
Consider a system in the state $|\Q\ket$ at time 
$t_{0}$ and evolving according to the time-dependent Schr\"odinger 
equation with Hamiltonian $\hat{H}(t)$. At this stage the state $|\Q\ket$ is a 
general many-body state and does not have to be the ground state of 
$\hat{H}(t_{0})$. We write 
\be
\hat{H}(t)=\hat{H}_{0}(t)+\hat{W}(t),
\ee
where 
\be
\hat{H}_{0}(t)=\int 
dx\,\hat{\q}^{\dag}(x)h(x,t)\hat{\q}(x)
\label{onebodyh}
\ee
is quadratic in the field operators $\hat{\q}(x)$ and  
$\hat{\q}^{\dag}(x)$ (here $x$ is a one-body quantum number like, 
e.g., the position-spin coordinate); this decomposition is arbitrary 
as $\hat{H}_{0}(t)$ can be any one-body operator.
The remaining term $\hat{W}(t)$ will often be a two-body operator
describing interactions between the particles, but can in fact be a
general $n$-body operator with any time-dependence. 
%In the following
%we will assume that it describes a two-particle interaction.
The basic quantities of the GFF 
are the $n$-th particle Green's functions 
\be
\addtolength{\fboxsep}{3pt}
\begin{split}
G_{n}&(1\ldots n;1'\ldots n') 
 \\
&= \frac{1}{i^n} \bra\Q|
\mathcal{T}\left\{
\hat{\q}_{H}(1)\ldots\hat{\q}_{H}(n)
\hat{\q}_{H}^{\dag}(n')\ldots\hat{\q}_{H}^{\dag}(1')
\right\}|\Q\ket,
\end{split}
\label{gn}
\ee
where $\mathcal{T}$ is the  time-ordering operator, 
the subscript ``$H$'' denotes the Heisenberg 
picture and the short-hand notation 
$1=(x_{1},t_{1})$, $1'=(x'_{1},t'_{1})$ etc. has been introduced. 
The $n$-body Green's function gives a complete description of $n$-body
correlations in a many-particle system.
If we write the evolution operator as 
$\hat{U}(t,t_{0})=\hat{U}_{0}(t,t_{0})\hat{F}(t,t_{0})$, 
where $\hat{U}_{0}$ is the 
non-interacting evolution operator, then $\hat{F}$ is the evolution 
operator in the interaction picture and fulfills
\be
i\frac{d}{dt}\hat{F}(t,t')=\hat{W}_{I}(t)\hat{F}(t,t')
\label{deff}
\ee 
with boundary condition
$\hat{F}(t,t)=1$  (the 
subscript ``$I$''  denotes the interaction picture). Consequently, we can rewrite $G_{n}$ 
(omitting its arguments) as
\beq
G_{n}
&=& \frac{1}{i^{n}}\bra\Q|
\hat{F}(t_{0},\inf)
\mathcal{T}\left\{e^{-i\int_{-\inf}^{\inf}d\bar{t}\,\hat{W}_{I}(\bar{t})}
\hat{\q}_{I}(1)\ldots\hat{\q}_{I}(n)\right.
\nonumber \\
&\times&\left.
\hat{\q}_{I}^{\dag}(n')\ldots\hat{\q}_{I}^{\dag}(1')
\right\}\hat{F}(-\inf,t_{0})|\Q\ket
\label{gn2}
\eeq
The Wick 
theorem applies to the $g_{m}$'s, i.e., the time-ordered product of field operators in 
the interaction picture {\em averaged over a non-interacting 
 state} 
$|\F_{0}\ket$.  
Thus to express $G_{n}$ in terms of the $g_{m}$'s Eq. (\ref{gn2}) 
needs to be further manipulated. Let us choose 
$|\F_{0}\ket$  as the ground state of 
$\hat{H}_{0}=\hat{H}_{0}(t_{0})$ and construct the 
Hamiltonian 
\be
\hat{H}^{\rm M}=\hat{H}_{0}+\hat{W}^{\rm M}
\label{matsham}
\ee
with ground state 
$|\Q\ket$.\cite{deg-note} According to the  Gell-Mann 
and Low theorem \cite{FetterWalecka,RungeGrossHeinonen} 
the state $|\Q\ket$ (for a more mathematical discussion
on the precise conditions see Refs. \onlinecite{nr.1989, bps.2010}) can be reached 
from $|\F_{0}\ket$ by adiabatically turning on the interaction 
$\hat{W}^{\rm M}$. Then,
\be
|\Q\ket=\hat{F}^{\rm M}(t_{0},-\inf)
|\F_{0}\ket,
\label{eq7}
\ee
where $\hat{F}^{\rm M}$ fulfills equation (\ref{deff}) with 
$\hat{W}\ra\hat{W}^{\rm M}e^{-\eta|t-t_{0}|}$ and $\eta$ an 
infinitesimal positive constant. Equation (\ref{eq7}) also implies 
that we can reach the state $|\Q\ket$ (up to a phase factor\cite{FetterWalecka,RungeGrossHeinonen}) 
with a backward propagation in time and hence
% Under the assumption that 
% an adiabatic switching on and off of the interaction $\hat{W}^{\rm M}$ 
% leaves the system in the same state $|\F_{0}\ket$, the state  
% $|\Q\ket$ can also be reached by a backward 
% propagation in time 
\be
|\Q\ket=e^{i\a}\hat{F}^{\rm M}(t_{0},\inf)
|\F_{0}\ket,
\ee
Exploiting these results  the Green's function $G_{n}$ in equation 
(\ref{gn2}) takes the form
\beq
G_{n}
&=&\frac{e^{-i\a}}{i^{n}}\bra\F_{0}^{+}|
\mathcal{T}\left\{e^{-i\int_{-\inf}^{\inf}d\bar{t}\,\hat{W}_{I}(\bar{t})}
\hat{\q}_{I}(1)\ldots\hat{\q}_{I}(n)\right.
\nonumber \\
&\times&\left.
\hat{\q}_{I}^{\dag}(n')\ldots\hat{\q}_{I}^{\dag}(1')
\right\}|\F_{0}^{-}\ket,
\label{gn3}
\eeq
where
\be
|\F_{0}^{\pm}\ket=\hat{F}(\pm\inf,t_{0})\hat{F}^{\rm M}(t_{0},\pm\inf)|\F_{0}\ket.
\ee
In most textbooks the interaction $\hat{W}(t)=\hat{W}$ is 
time-independent and $|\Q\ket=|\Q_{0}\ket$ is the 
ground state of $\hat{H}=\hat{H}(t_{0})$; then  
$|\F_{0}^{\pm}\ket=|\F_{0}\ket$ since  $\hat{W}^{\rm M}=\hat{W}$. In 
this case we can expand 
the exponent in equation (\ref{gn3}) in powers of 
$\hat{W}_{I}$, express $G_{n}$ in 
terms of the noninteracting $m$-particle Green's functions $g_{m}$'s 
with $m\geq n$ and in turn the $g_{m}$'s 
in terms of $g_{1}$ using the Wick  theorem.\cite{phase}
To the contrary if the 
interaction is time-dependent and/or
$|\Q\ket\neq|\Q_{0}\ket$, the 
standard Wick theorem is of no use to calculate 
$G_{n}$. Our aim is therefore to lift these restrictions in a generalized version of
Wick's theorem which can be applied to general initial states for both equilibrium and
non-equilibrium systems.

\begin{figure}
\begin{center}
\includegraphics[width=7.cm]{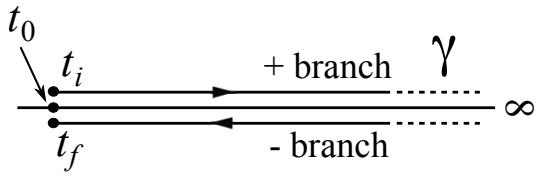}
\caption{The original Keldysh contour. The contour 
time $z=t_{\pm}$ lies on the $\pm$ branch at a distance $t$ from the 
origin. The 
end-points are $t_{i}=t_{0-}$ and $t_{f}=t_{0+}$.}
\label{contour}
\end{center}
\end{figure}

\section{Generalized Wick Theorem}
Reading the time-arguments in equation (\ref{gn2}) from right to left it is 
natural to design the Keldysh contour\cite{k.1965} $\g$  of Fig.
\ref{contour} and define the 
generalized Green's function 
\beq
G_{n} &=&
\frac{1}{i^{n}} \bra\Q|
\mathcal{T}_{\g}\left\{e^{-i\int_{\g} d\bar{z}\,\hat{W}_{I}(\bar{z})}
\hat{\q}_{I}(1)\ldots\hat{\q}_{I}(n)\right.
\nonumber \\
&\times&
\left.\hat{\q}_{I}^{\dag}(n')\ldots\hat{\q}_{I}^{\dag}(1')
\right\}|\Q\ket,
\label{greenfunc}
\eeq
with $\mathcal{T}_{\g}$ the contour-ordering operator and 
$1=(x_{1},z_{1})$,  $1'=(x'_{1},z'_{1})$ etc. collective indices with 
times $z$'s on the contour. The new $G_{n}$ coincides with the time-ordered 
one of Eq. (\ref{gn2}) when all contour-times lie on the 
upper branch. Expanding the exponent in powers of $\hat{W_{I}}$, the 
Keldysh $G_{n}$ can be written as the sum of integrals over 
non-interacting Keldysh 
Green's functions
\be
\addtolength{\fboxsep}{3pt}
\begin{split}
g_{m}&(1\ldots m;1'\ldots m') 
 \\
&=\frac{1}{i^{m}}\bra\Q|
\mathcal{T}_{\g}\left\{
\hat{\q}_{I}(1)\ldots\hat{\q}_{I}(m)\hat{\q}_{I}^{\dag}(m')\ldots\hat{\q}_{I}^{\dag}(1')
\right\}|\Q\ket
\end{split}
\label{gmkel}
\ee
where $m>n$. For these functions a generalized Wick theorem will now be derived.

From the equations of motion of the field operators 
$\hat{\q}_{I}$Ê
and $\hat{\q}_{I}^{\dag}$ it follows immediately that 
the $g_{m}$ in Eq. (\ref{gmkel}) fulfill the non-interacting 
MSH\cite{ms.1959}
\beq
[i\de_{z_{k}}-h(k)]g_{m}=
\sum_{j=1}^{m}(\pm)^{k+j}\d(k,j')g_{m-1}(\breve{k};\breve{j} '),
\nonumber \\ 
\,[-i\de_{z'_{j}}-h(j')]g_{m}=\sum_{k=1}^{m}(\pm)^{k+j}\d(k,j')g_{m-1}(\breve{k};\breve{j}'),
\label{ni-msh}
\eeq
where $\d(k,j')=\d(x_{k},x'_{j})\d(z_{k},z'_{j})$, $h$ is the one-body
Hamiltonian of Eq. (\ref{onebodyh}) and the upper/lower sign refers 
to bosons/fermions. Here and in the 
following the symbol `` $\breve{}$ '' over the indices specifies the 
missing indices of $g_{m}$.
Thus for instance $g_{3}(\breve{1};\breve{2}')=g_{3}(23;1'3')$.
It is worth noting that Eqs. (\ref{ni-msh}) 
constitute a system of recursive 
relations since $g_{m}$ can be calculated from the sole knowledge of $g_{m-1}$. 
The solution for $g_m$ depends, of course, on the boundary conditions  that we impose.
The boundary conditions for $g_m$ follow directly from its definition 
(\ref{gmkel})
\be
\addtolength{\fboxsep}{3pt}
\begin{split}
\G_{m}(x_{1}\ldots x_{m}&;x'_{1}\ldots 
x'_{m})
\\
&=(\pm i)^{m}\lim_{z_{k},z'_{j}\to t_{i}}g_{m}(1\ldots m;1'\ldots m')
\end{split}
\label{bc}
\ee
with 
$z_{1}<\ldots <z_{m}<z'_{m}<\ldots<z'_{1}$ (from now on 
$\lim_{z_{k},z'_{j}\to t_{i}}$ will always be taken in such order)
where $t_i$ is the initial time of Fig. \ref{contour} and
\be
\addtolength{\fboxsep}{3pt}
\begin{split}
\G_{m}(x_{1}\ldots& x_{m};x'_{1}\ldots 
x'_{m})
\\
&=\bra\Q|\hat{\q}^{\dag}(x'_{1})\ldots\hat{\q}^{\dag}(x'_{m})\hat{\q}(x_{m})
\ldots\hat{\q}(x_{1})|\Q\ket
\end{split}
\label{GammaM}
\ee
is the initial $m$-body density matrix describing the initial $m$-body correlations. 
One can readily verify by expanding along a row or a column that the permanent/determinant
\beq
g_{m}(1\ldots m;1'\ldots m')&=&\left|\begin{array}{ccc}
g(1;1') & \ldots & g(1;m') \\
\vdots & & \vdots \\
g(m;1') & \ldots & g(m;m')
\end{array}\right|_{\pm}
\nonumber \\
&\equiv &
|g|_{m}
\label{standard-wick}
\eeq
of one-particle Green's functions $g\equiv g_{1}$ is a solution of 
Eq. (\ref{ni-msh}). However, 
in general the solution (\ref{standard-wick}) 
will not satisfy the boundary conditions (\ref{bc}). 
This will only happen when the 
$m$-particle density matrix $\G_m$ of equation (\ref{GammaM})
is averaged over a non-interacting state $|\Q\ket=|\F_{0}\ket$.
Note that this result constitutes a {\em one-line proof of the 
standard Wick theorem} which, in our formulation, amounts to solving 
a boundary problem for the MSH.

\begin{figure}
\begin{center}
\includegraphics[width=8.cm]{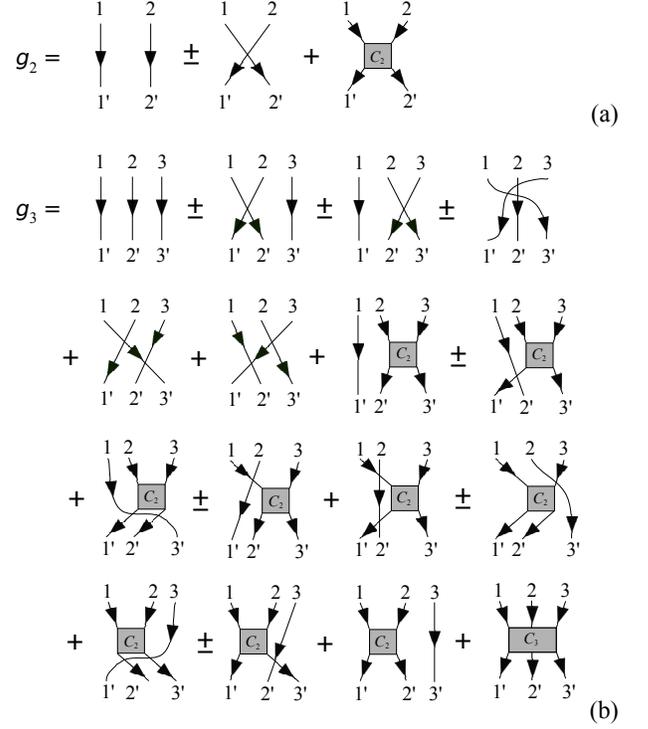}
\caption{Diagrammatic representation of $g_{2}$ (a) and $g_{3}$ (b).}
\label{g2g3}
\end{center}
\end{figure}

For arbitrary initial states $|\Q\ket$ the particular solution 
(\ref{standard-wick}) must  
be supplied with the additional solution $\tilde{g}_{m}$ of the homogeneous 
equations
\be
[i\de_{z_{k}}-h(k)]\tilde{g}_{m}=0
\label{homogen}
\ee 
and its adjoint to 
satisfy the boundary conditions. For illustrative purposes we 
first consider the examples of $g_{2}$ and $g_{3}$. The most general 
solution of Eqs. (\ref{ni-msh}) for $m=2$ 
reads 
\be
g_{2}=|g|_{2}+\tilde{g}_{2}. 
\ee
Being a homogeneous 
solution, $\tilde{g}_{2}=g_{2}-|g|_{2}$ is not discontinuous when a contour-time 
$z_{k}$ passes through $z'_{j}$, which implies that the discontinuity of $g_{2}$ is 
compensated by an identical discontinuity in $|g|_{2}$. The 
initial value of $\tilde{g}_{2}$ is fixed by the boundary conditions (\ref{bc})
for $g_{2}$ and $g$ and reads 
\be
\lim_{z_{k},z'_{j}\to 
t_{i}}\tilde{g}_{2}(12;1'2')=(\mp 
i)^{2}C_{2}(x_{1}x_{2};x'_{1}x'_{2})
\label{bc2}
\ee 
with the correlation function
\be
C_{2}\equiv
\G_{2}-|\G|_{2}
\ee
and $\G\equiv\G_{1}$ the one-particle density matrix.
The limit in Eq. (\ref{bc2}) is now independent of the time-ordering in which the limit
is taken. 
The same applies to all functions $\tilde{g}_{m}$ defined by Eq. 
(\ref{homogen}), i.e.,
their limit is defined independent of the time-ordering taken.
We can express $\tilde{g}_{2}$ in terms 
of $g$ and $C_{2}$ as follows. 
The spectral function on the contour 
\beq
A(1;1')&\equiv&i\left[g^{>}(1;1')-g^{<}(1;1')\right]
\nonumber \\
&=& \bra\Q|\hat{\q}_{I}(1)\hat{\q}^{\dag}_{I}(1')\pm 
\hat{\q}^{\dag}_{I}(1')\hat{\q}_{I}(1)|\Q\ket
\eeq
satisfies the 
homogeneous equations 
\be
[i\de_{z_{1}}-h(1)]A(1;1')=[-i\de_{z'_{1}}-h(1')]A(1;1')=0,
\label{spectraleqn}
\ee (hence it 
has no discontinuity 
when $z_{1}$ passes through $z'_{1}$) and equals $\d(x_{1},x_{2})$ for 
$z_{1}=z'_{1}$. If we define 
\be
\d_{-}(z)\equiv\d(z,t_{i})-\d(z,t_{f}),
\ee 
where $t_{i}$ and $t_{f}$ are the end-points of the Keldysh contour 
of Fig. \ref{contour},
we can write 
\beq
A(1,x'_{1}t_{i}) &=&i\left[ g(1,x'_{1}t_{i})-g(1,x'_{1}t_{f}) \right]
\nonumber \\
&=&i \int_{\g}d\bar{z}\,g(1,x'_{1}\bar{z})\d_{-}(\bar{z})
\eeq
and similarly \cite{notedelta}
\be
A(x_{1}t_{i},1') =-i\int_{\g}d\bar{z}\,\d_{-}(\bar{z})g(x_{1}\bar{z},1').
\ee
Therefore, introducing
the time-local correlation function on the contour 
\beq
C_{2}(12;1'2') &\equiv& \d_{-}(z_{1})\d_{-}(z_{2}) 
\nonumber \\
&\times& C_{2}(x_{1}x_{2};x'_{1}x'_{2})\d_{-}(z'_{1})\d_{-}(z'_{2}) 
\eeq
it is immediate to realize that
\beq
\tilde{g}_{2}(12;1'2')&=&(\mp i)^{2}\int_{\g}
d\bar{1}d\bar{2}d\bar{1'}d\bar{2'}
g(1;\bar{1})g(2;\bar{2})
\nonumber \\
&\times&C_{2}(\bar{1}\bar{2};\bar{1'}\bar{2'})
g(\bar{1'};1')g(\bar{2'};2') 
\eeq
provides an explicit solution to the initial value problem for 
$g_{2}$, i.e., it satisfies Eqs. (\ref{homogen}) and (\ref{bc2}). 
Here and in the following $\int_{\g}d1=\int dx_{1}\int_{\g}dz_{1}$.
The diagrammatic representation of the generalized Wick 
theorem for $g_{2}$ is displayed  in Fig. \ref{g2g3}(a); it is worth noting that for a non-interacting 
initial state $C_{2}=0$  and the standard 
Wick theorem is recovered.

Inserting $g_{2}=|g|_{2}+\tilde{g}_{2}$ into Eq. 
(\ref{ni-msh}) we find in a similar way that the general solution for $g_{3}$ is
\be
g_{3}=|g|_{3}+\sum_{k,j=1}^{3}(\pm)^{k+j}g(k;j')\tilde{g}_{2}(\breve{k};\breve{j}')
+\tilde{g}_{3},
\label{g3}
\ee
where the homogeneous solution $\tilde{g}_{3}$ is fixed by the 
boundary conditions for $g_{3}$, $g_{2}$ and $g$: 
\be
\lim_{z_{k},z'_{j}\to 
t_{i}}\tilde{g}_{3}(123;1'2'3')=(\mp 
i)^{3}C_{3}(x_{1}x_{2}x_{3};x'_{1}x'_{2}x'_{3}),
\ee 
with 
\be
C_{3}=\G_{3}-\sum_{k,j=1}^{3}(\pm)^{k+j}\G(x_{k};x'_{j})
C_{2}(\breve{x}_{k};\breve{x}'_{j})-|\G|_{3}. 
\ee
Using the (anti)-symmetry of $\tilde{g}_2$ and $\tilde{g}_3$ one can readily check that
$g_3$ in Eq. (\ref{g3}) is properly (anti)-symmetric in the primed and the unprimed variables.
The second term in Eq. (\ref{g3}) has exactly the same structure of a 
generalized Laplace expansion for permanents/determinants. As we 
shall see this is a general feature of all terms of the expansion 
of $g_{m}$.
Similarly to the two-particle case we can define the time-local density matrix function 
on the contour and write 
\beq
\tilde{g}_{3}(123;1'2'3')&=&(\mp i)^{3}\int_{\g} \prod_{k=1}^{3}
d\bar{k}\,g(k;\bar{k})
\nonumber \\
&\times&
C_{3}(\bar{1}\bar{2}\bar{3};\bar{1'}\bar{2}'\bar{3}')
\prod_{j=1}^{3}d\bar{j}'g(\bar{j}';j').
\nonumber
\eeq
The diagrammatic representation of the generalized Wick theorem for 
$g_{3}$ is displayed in Fig. \ref{g2g3}(b).  
For instance, the second diagram of the last line in Fig. \ref{g2g3}(b)
gives the contribution $(\pm)^{3+2} g(3,2') \tilde{g}_2 (12;1'3')$ to 
Eq. (\ref{g3}).
Note that the sign of a diagram is given by $(\pm)^{n_c}$ where
$n_c$ is the number of crossings of Green's function lines. Alternatively it is given
by $(\pm)^{n_c}$ where $n_c$ is the number of transpositions needed to reorder
the labels $k \breve{k}$ and $j' \breve{j}'$ in Eq. (\ref{g3}) to 
$123$ and $1'2'3'$, as is readily checked for our example. 

\begin{figure}
\begin{center}
\includegraphics[width=8.cm]{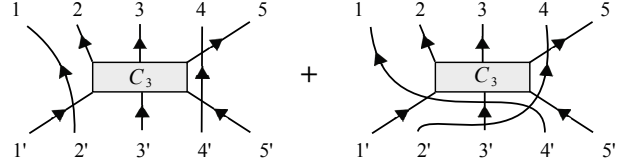}
\caption{Diagrammatic representation of the terms of the expansion 
(\ref{gen-wick0}) for $m=5$ and $l=2$ with 
$P=(1,4)$ 
and $Q=(2,4)$ corresponding to Eq. (\ref{c5}).}
\label{general-wick}
\end{center}
\end{figure}

At this point it is not difficult to guess the solution for $g_{m}$. 
Let us introduce the collective ordered indices $K=(k_{1}\ldots 
k_{l})$, with $k_{1}<\ldots<k_{l}$ and $X_{K}=(x_{k_{1}}\ldots x_{k_{l}})$,  
and define 
\be
C_{l}(K;J')=\left(\prod_{\a=1}^{l}\d_{-}(z_{k_{\a}})\right)
C_{l}(X_{K};X'_{J}) \left(\prod_{\b=1}^{l}\d_{-}(z'_{j_{\b}})\right)
\ee
where $C_{l}(X_{K};X'_{J})$ is defined by the recursive relations
\beq
C_{l}(X_{K};X'_{J})&=&\G_{l}(X_{K};X'_{J})-\sum_{n=1}^{l-2}\sum_{PQ}(\pm)^{|P+Q|}
\nonumber \\
&\times&
|\G|_{n}(X_{P};X'_{Q})C_{n-l}(\breve{X}_{P};\breve{X}'_{Q})
\nonumber \\
&-&|\G|_{l}(X_{K};X'_{J})
\label{tildeGl}
\eeq
with $C_{2}\equiv \G_{2}-|\G|_{2}$. In equation (\ref{tildeGl}) the sum runs over all ordered 
sequences $Q=(q_{1}\ldots q_{n})$, $P=(p_{1}\ldots p_{n})$ with 
indices between $1$ and $l$, and the 
sign is determined by 
\be
|P+Q| \equiv \sum_{\a=1}^{n}(q_{\a}+p_{\a}). 
\ee
For example, let $l=5$ and $n=2$ with $P=(1,4)$  and $Q=(2,4)$ and 
hence $\breve{P}=(2,3,5)$ and $\breve{Q}=(1,3,5)$. We then 
have $X_{P}=(x_{1},x_{4})$, $X'_{Q}=(x'_{2},x'_{4})$ and the 
complementary collective coordinates
$\breve{X}_{P}=(x_2 ,x_3 ,x_5)$,
$\breve{X}'_{Q}=(x_1', x_3', x_5')$. The corresponding term 
for $C_5$ under the summation sign in (\ref{tildeGl}) is   
given by
\beq
\lefteqn{ (\pm)^{1+4+2+4} | \Gamma|_{2} (x_1 x_4; 
x_2' x_4')  C_{3} (x_2 x_3 x_5; x_1' x_3' x_5')  } 
\nonumber \\
&=& \!\!\!\pm \left|
\begin{array}{cc}
 \Gamma (x_1;x_2') &  \Gamma (x_1;x_4')    \\
 \Gamma (x_4;x_2') &  \Gamma (x_4;x_4')
\end{array}
\right|_{\pm}   C_{3} (x_2, x_3 ,x_5; x_1', x_3', 
x_5'). 
\nonumber \\
\label{c5}
\eeq
We will now prove the generalized Wick theorem:

- {\em Theorem}: The solution of the MSH of Eqs. (\ref{ni-msh}) with 
the boundary conditions  (\ref{bc}) is
\be
g_{m}=|g|_{m}+\sum_{l=1}^{m-2}g_{m}^{(l)}+\tilde{g}_{m}
\label{gen-wick0}
\ee
where we defined
\be
g_{m}^{(l)}(K;J') \equiv
\sum_{PQ}(\pm)^{|P+Q|}|g|_{l}(P;Q')
\tilde{g}_{m-l}(\breve{P};\breve{Q}')
\label{gen-wick}
\ee
and, in complete analogy with $\tilde{g}_{2}$ and $\tilde{g}_{3}$,
\beq
 \tilde{g}_{l}(K;J') &=&  
 (\mp i)^{l}\int_{\g} \prod_{\a=1}^{l}d\bar{k}_{\a}\,g(k_{\a};\bar{k}_{\a})
 \nonumber \\
 &\times&
C_{l}(\bar{K};\bar{J'})\prod_{\b=1}^{l} d\bar{j}'_{\b}\,
g(\bar{j}'_{\b};j'_{\b})  .
\label{gen-wick2}
\eeq
Equations (\ref{gen-wick0})-(\ref{gen-wick2}) give a complete 
solution of the initial value problem  
for the MSH for both equilibrium and nonequilibrium systems. The only 
input is the  
initial $m$-body correlations incorporated in $\G_m$.
%\be
%\tilde{g}_{l}(K;J')=(\mp i)^{l}\int_{\g} (\prod_{\a=1}^{l}g(k_{\a};\bar{k}_{\a}))
%\tilde{\G}^{\d}_{l}(\bar{K};\bar{J'})(\prod_{\b=1}^{l} 
%g(\bar{j}'_{\b};j'_{\b}))  \nonumber 
%\ee
Let us make some additional remarks. 
% In equation (\ref{gen-wick}) we implicitly defined $|g|_{0}=1$ so that 
% $g_{m}^{(m)}=\tilde{g}_{m}$.
The function $g_{m}^{(l)}$ is the sum of all possible diagrams with a single 
correlation block $C_{m-l}$ in which $m-l$ Green's function lines enter and leave, multiplied
by $l$ separate Green's function lines. 
For instance, for $m=5$ and $l=2$ the term of the sum in Eq. (\ref{gen-wick0}) with 
$P=(1,4)$ 
and $Q=(2,4)$ has the  diagrammatic representation in Fig. 
\ref{general-wick} and corresponds to Eq. (\ref{c5}).
The sign of the diagrams is $(\pm)^{n_{c}}$ where  $n_{c}$ equals the 
number of crossings (in Fig. \ref{general-wick} $n_{c}=3$ for 
the first diagram and $n_{c}=6$ for the second diagram).
Although the general structure of $g_{m}$ was 
outlined in Refs. \onlinecite{h.1975,d.1984} our theorem contains a precise statement on how to 
construct it. 
When $|\Q\ket$ is a non-interacting product state
 the recursive relations in Eq. 
(\ref{tildeGl}) yield $C_{l}=0$, hence $\tilde{g}_{l}=0$ and 
consequently $g_{m}=|g|_{m}$, i.e., we recover the standard Wick 
theorem,\cite{disconnected} in agreement with the previous discussion.
The generalized Wick theorem superseeds all common 
limitations of the standard 
GFF, i.e., the non-degeneracy of the ground state, the assumptions of 
the Gell-Mann and Low theorem  and the requirement of a 
time-independent Hamiltonian for times $t\ra \pm \inf$.
Finally we would like to point out that 
Eq. (\ref{gen-wick0}) has the 
structure of the permanent/determinant of 
the sum of two matrices $A$ and $B$. Indeed according to
the Laplace formula 
\beq
|A+B|_{m} &=& |A|_{m} + \sum_{l=1}^{m-1} \sum_{PQ} 
(\pm)^{|P+Q|}
\nonumber \\ 
&\times&
|A|_{l}(P;Q) |B|_{m-l} (\breve{P};\breve{Q})  
+  |B|_{m}  
\label{sumdet}
\eeq
where $|A|_{l}(P;Q)$ is the permanent/determinant of the $l\times 
l$ matrix obtained with the rows $P$ and the 
columns $Q$ of the matrix $A$. In the special case $l=m$ we have 
$P=Q=(1\ldots m)$ and hence
$|A|_{m}(P;Q)=|A|_{m}$. The same notation has been used for 
the matrix $B$.
With the identification $A_{kj}=g(k;j')$ and $|B|_{m-l} 
(\breve{P};\breve{Q}) = \tilde{g}_{m-l} 
(\breve{P};\breve{Q}')$   
for $l=1 \ldots m-2$ and the definition $\tilde{g}_1 \equiv 0$
Eqs. (\ref{gen-wick0}) and (\ref{sumdet})  become identical.
We can thus symbolically write the 
generalized Wick theorem  as 
\be
g_{m} = |g + \tilde{g}|_{m}
\ee
whose precise meaning  is given by Eq. (\ref{gen-wick0}).

- {\em Proof}: The proof consists of two steps. We first prove that $g_{m}$ satisfies 
the MSH. For this purpose it is enough to prove the MSH for each 
$g_{m}^{(l)}$ since, as already noticed, the MSH is obviously 
satisfied by $|g|_{m}$. We will consider only the first of equations 
(\ref{ni-msh}) as the same reasoning applies to the second.
 If we act with $(i\de_{z_{k}}-h(k))$ 
on a particular diagram of $g_{m}^{(l)}$ the result is zero if 
$k$ is an ingoing line to a $C_{l}$-block. 
This follows immediately from the definition (\ref{gen-wick2})
and Eq. (\ref{spectraleqn}).
To the contrary, 
if the diagram contains a free line
$g(k;j')$ the result is $(\pm)^{k+j}\d(k,j')\times$[a diagram for 
$\tilde{g}_{m-1}^{(l)}(\breve{k};\breve{j}')$]. Thus,  
$(i\de_{z_{k}}-h(k))\sum[$all diagrams in $g_{m}^{(l)}$ 
containing 
$g(k;j')]=(\pm)^{k+j}\d(k,j')\tilde{g}_{m-1}^{(l)}(\breve{k};\breve{j}')$. The MSH 
for $g_{m}^{(l)}$ follows from this relation when summing over all 
$j$. Taking into account that $(i\de_{z_{k}}-h(k))\tilde{g}_{m}=0$ we 
conclude that $g_{m}$ satisfies the MSH.

Next we prove that $g_{m}$ has the correct boundary conditions (\ref{bc}). 
In Eq. (\ref{gen-wick0}) we take the limit
$z_{k},z'_{j}\to t_{i}$. Since $|g|_{l}\to (\mp i)^{l}|\G|_{l}$ 
and $\tilde{g}_{m-l}\to (\mp i)^{m-l} C_{m-l}$ we find
\be
\lim_{z_{k},z'_{j}\to t_{i}} (\pm i)^m g_{m}=|\G|_{m}+\sum_{l=1}^{m-2}
\sum_{PQ}(\pm)^{|P+Q|}|\G|_{l}C_{m-l}+C_{m}.
\ee
Writing the last term $C_{m}$ as in equation (\ref{tildeGl})
we see by inspection that all terms of the sum cancel out and the 
r.h.s. reduces to $\G_{m}$. 
We therefore exactly satisfy the boundary condition (\ref{bc}). This 
concludes the proof $\blacksquare$

\begin{figure}
\begin{center}
\includegraphics[width=8.cm]{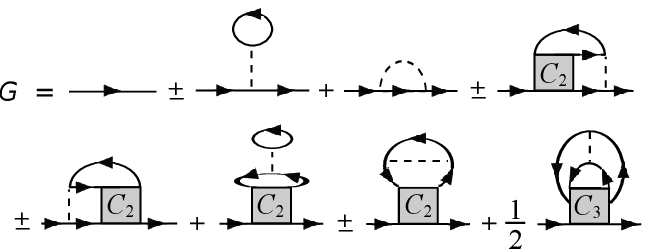}
\caption{First-order expansion of $G$ for a two-body interaction. The 
last three diagrams vanish since both lines enter a 
$C_{l}$-block and hence the internal time-integration 
can be reduced to a point. This is a completely general feature. }
\label{self-energy}
\end{center}
\end{figure}

\section{Generalized Green's function formalism} 
With our generalized Wick theorem we can now directly expand 
the one-particle Green's function $G\equiv G_{1}$ in powers 
of the interaction $\hat{W}_I$, see Eq. (\ref{greenfunc}).
In what follows we restrict the analysis to interactions which are 
represented by two-body operators like, e.g., the Coulomb interaction.
From Eq. (\ref{greenfunc}) the interacting Green's function is then given by
\beq
G(a,b) &=& \sum_{k=0}^\infty \frac{1}{k!} (\frac{i}{2})^k \int_\gamma w(1,1')\ldots w(k,k')
\nonumber \\
&\times&g_{2k+1} (a,1,1' \ldots k,k';b,1^+,1'^+ \ldots k^+,k'^+)
\nonumber \\
\label{Gexpansion}
\eeq
in which $w(j,j')$ is the two-body interaction and we integrate over 
all space-time coordinates $j$ and $j'$ for $j=1 \ldots k$. 
By inserting Eq. (\ref{gen-wick0}) into Eq. 
(\ref{Gexpansion}) we obtain a diagrammatic 
expansion of the interacting Green's function. For example, to first 
order in the interaction we only need the $g_3$ 
of Fig. \ref{g2g3}. If we then take into account that the 
disconnected pieces vanish 
we find that to first order in the interaction the Green's function is 
given by the connected diagrams displayed in Fig. 
\ref{self-energy}.  
In general the expansion of $G$ in powers of $\hat{W}_I$ leads to a 
diagrammatic series which starts end ends with a $g$-line. The kernel 
of this expansion is therefore the {\em  reducible} self-energy which 
we denote by
$\S^{(r)}_{\rm tot}$.
In the case of general initial states the 
$\S^{(r)}_{\rm tot}$-diagrams contain at most one correlation block 
and either begin and 
end with an interaction line 
(we call their sum $\S^{(r)}$) or begin/end with an interaction line 
and end/begin with a correlation block (we call their sum 
left/right reducible self-energy  $\S^{(r)}_{L,R}$).\cite{d.1984}
This means that the general structure of the diagrammatic expansion is 
\be
G(1;2) =g(1;2)+\int_{\g} g(1;\bar{1})\S^{(r)}_{\rm tot}(\bar{1};\bar{2}) 
g(\bar{2};2)
\label{reddeq}
\ee
where $\S^{(r)}_{\rm tot}=\S^{(r)}+\S_{L}^{(r)}+\S_{R}^{(r)}$
and \cite{keldprop}
\be
\S_{L}^{(r)}(1;2) = \S_{L}^{(r)}(1;x_{2})\d_{-}(z_{2}) ,
\ee
\be
\S_{R}^{(r)}(1;2) = \d_{-}(z_{1})\S_{R}^{(r)}(x_{1};2).
\ee
Thus the self-energy is
modified by the addition of a left/right reducible self-energy which is 
non-zero only if its right/left time-argument is one of the 
end-points of the Keldysh contour. The diagrammatic expansion of 
these new self-energies  is very similar to
that of the standard GFF, as it is exemplified in 
Fig. \ref{self-energy}. The only extra ingredient is the appearance of 
the $C_{m}$-blocks which describe the initial $m$-body correlations.

\section{Relation to the imaginary time formalism and existence of a Dyson equation}

\begin{figure}[tbp]
\begin{center}
\includegraphics[width=7.cm]{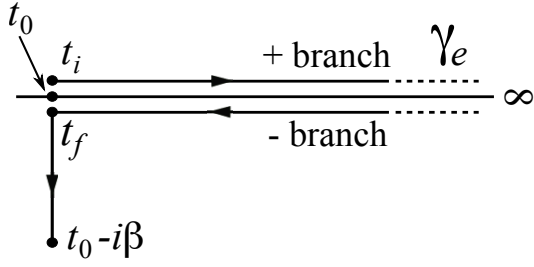}
\caption{Extended Keldysh contour with an imaginary track between 
$t_{0}$ and $t_{0}-i\b$. The contour 
time $z=t_{\pm}$ lies on the $\pm$ branch at a distance $t$ from the 
origin and $z=t_{0}-i\t$ lies on the imaginary track. The 
end-points are $t_{i}=t_{0-}$ and $t_{f}=t_{0}-i\b$.}
\label{contour2}
\end{center}
\end{figure}

We next discuss the 
relation between the above results and the imaginary-time 
formalism\cite{d.1984,w.1991} and in doing so will also generalize some
recent work by Garny and M\"uller. \cite{garny.2009}
The average over the state $|\Q\ket$ can 
also be computed by tracing the field operators with 
$
\hat{\r}(\b)=e^{-\b \hat{H}^{\rm M}}/{\rm Tr}[e^{-\b \hat{H}^{\rm 
M}}]
$
where $\beta$ is a real positive number (inverse temperature)
 and then letting $\b\to\inf$.
We remind the reader that $\hat{H}^{\rm M}$ is the Hamiltonian with 
$|\Q\ket$ as ground state, see Eq. (\ref{matsham}), and can be a general $n$-body
operator. This is discussed
in more detail in Refs. \onlinecite{d.1984,w.1991,mr.1999}.
Then, 
\begin{widetext}
\be
G_{n}(1\ldots n;1'\ldots n') = \frac{1}{i^{n}} \frac{ {\rm Tr}[e^{-\b\hat{H}_{0}}
\mathcal{T}_{\g_{e}}\{e^{-i\int_{\g_{e}} 
d\bar{z}\,\hat{W}_{I}(\bar{z})}
\hat{\q}_{I}(1)\ldots\hat{\q}_{I}(n) 
\hat{\q}_{I}^{\dag}(n')\ldots\hat{\q}_{I}^{\dag}(1')\}]
}
{ {\rm Tr}[e^{-\b \hat{H}^{\rm 
M}}] }
\ee
\end{widetext}
where $\g_{e}$ is the extended Keldysh contour of Fig. \ref{contour2} 
and $\hat{W}(z)=\hat{W}^{\rm M}$ for $z$  on the imaginary 
track of the extended contour.\cite{stefalm.2004,TDDFTbook}
These new Green's functions coincide with the old ones when all 
contour-times lie on the original Keldysh contour $\g$. The advantage of this formulation is that 
the corresponding new non-interacting 
Green's functions can be expanded using the standard Wick theorem, 
$g_{0,m}=|g_{0}|_{m}$, since both $g_{0,m}$ and $g_{0}$ obey the 
Kubo-Martin-Schwinger boundary 
conditions,\cite{kms.1957-59,ms.1959} i.e., they are periodic/antiperiodic in 
the contour-time arguments. Note that $g_{0,m}\neq g_{m}$: the 
former is averaged with $\hat{\r}_{0}(\b)=e^{-\b \hat{H}_{0}}/
{\rm Tr}[e^{-\b \hat{H}_{0}}]$ while the latter with $\hat{\r}(\b)$.
The new Green's function with both arguments on $\g$ obeys the 
equations of motion\cite{d.1984,dahlen.2007,TDDFTbook,stefalm.2004} 
(integral over barred variables is understood)
\beq
[i\de_{z_{1}}-h(1)]G(1,2)&=&\d(1,2)+\int_{\g} \S_{e}(1;\bar{1})G(\bar{1};2)
\nonumber \\
&-&i\int_{0}^{\b}\S^{\rceil}_{e}(1;\bar{1})G^{\lceil}(\bar{1};2),
\label{eomext}
\eeq
and its adjoint, where $\S_{e}$ is the standard self-energy, i.e., 
the same as in Wagner work,\cite{w.1991} with 
internal integrals over all space and contour-times on $\g_{e}$.
 In Eq. (\ref{eomext}) 
the symbol ``$\rceil$'' specifies that the first contour argument 
lies on $\g$ and the second on the imaginary track; the opposite is 
specified by ``$\lceil$''.\cite{TDDFTbook,stefalm.2004} We will now derive an 
elegant relation between $\S_{e}$ and the reducible self-energy 
$\S^{(r)}_{\rm tot}$, relation which will be used to prove the 
existence of a Dyson equation on $\g$ for general initial states.
From the Langreth rules on $\g_{e}$ 
\cite{stefalm.2004} we have 
\beq
G^{\lceil}(1;2) &=&-i\int d\bar{x}\,G^{\rm 
M}(1;\bar{x} t_{0})G^{\rm A}(\bar{x}t_{0};2)
\nonumber \\
&+&[G^{\rm M}\star 
\S_{e}^{\lceil}\cdot G^{\rm A}](1;2).
\eeq
In this equation $
G^{\rm R/A}(1;2)=\pm \theta(\pm t_{1}\mp 
t_{2})[G^{>}(1;2)-G^{<}(1;2)]
$
is the retarded/advanced Green's function, $G^{\rm M}$ is the Matsubara 
Green's function with both arguments on the imaginary track, and we used 
the short-hand notation ``$\star$'' for convolutions between $t_{0}$ 
and $t_{0}-i\b$ and ``$\cdot$'' for convolutions between $t_{0}$ and 
$\inf$. Inserting this result into the last term of Eq. 
(\ref{eomext}) and taking 
into account that for any function $f(xt_{\pm})=f(xt)$ it holds  
\be 
\int_{\g}f(\bar{1})G(\bar{1};2)=
\int_{t_{0}}^{\inf}d\bar{x}d\bar{t}f(\bar{x}\bar{t})G^{\rm A}(\bar{x}\bar{t},2),
\ee
we find
\beq
&&[i\de_{z_{1}}-h(1)]G(1;2)=\d(1;2)
\nonumber \\
&&+\left[\left(
\S_{e}+\S_{e}^{\rceil}\star G^{\rm 
M}\star\S^{\lceil}_{e}+\S_{e,L}\right)\cdot G\right](1;2),
\label{eq2}
\eeq
where
\be
\S_{e,L}(1;2)=-i[\S_{e}^{\rceil}\star G^{\rm 
M}](1;x_{2}t_{0})\d_{-}(z_{2}).
\ee
Similarly for the adjoint equation we have
\beq
&&[-i\de_{z_{2}}-h(2)]G(1;2)=\d(1;2)
\nonumber \\
&&+\left[G\cdot\left(
\S_{e}+\S_{e}^{\rceil}\star G^{\rm 
M}\star\S^{\lceil}_{e}+\S_{e,R}\right)\right](1;2),
\label{eq3}
\eeq
with
\be
\S_{e,R}(1;2)=\d_{-}(z_{1})i[G^{\rm M}\star 
\S_{e}^{\lceil}](x_{1}t_{0};2).
\ee
Given the self-energy and the Green's function with arguments on the 
imaginary track we can regard Eqs.  (\ref{eq2}) and (\ref{eq3}) as the 
equations of motion for $G$ with arguments on $\g$. To integrate 
these equations we cannot use the non-interacting Green's function 
$g_{0}$
which satisfies the Kubo-Martin-Schwinger relations as 
the point $t_{0}-i\b$ does not belong to $\g$. However we can use 
the non-interacting Green's function $g$ of Eq. (\ref{gmkel}) since 
it satisfies
\be
\lim_{z_{1},z_{2}\ra t_{i}}g(1;2)=
\lim_{z_{1},z_{2}\ra t_{i}}G(1;2)
=
\G(x_{1};x_{2}).
\ee
 Thus if we define the total self-energy as 
\be
\S_{\rm tot}=\S_{e}+\S_{e}^{\rceil}\star G^{\rm 
M}\star\S^{\lceil}_{e}+\S_{e,L}+\S_{e,R}
\ee
we can write the following Dyson equation on $\g$ (integral over barred 
variables is understood)
\be
G(1;2)=g(1;2)+\int_{\g}g(1;\bar{1})\S_{\rm tot}(\bar{1};\bar{2}) 
G(\bar{2},2).
\label{eq4}
\ee
Comparing now this equation with (\ref{reddeq}) we deduce an 
important relation between the reducible self-energy $\S^{(r)}_{\rm 
tot}$ written in terms of correlation blocks and the irreducible 
self-energy $\S_{\rm tot}$ written in terms of integral over the 
imaginary track
\be
\S^{(r)}_{\rm tot}=\S_{\rm tot}+\S_{\rm tot}g\S_{\rm tot}+\S_{\rm tot}g\S_{\rm 
tot}g\S_{\rm tot}+\ldots
\label{eq5}
\ee
This relation constitute a bridge between two different and complementary 
approaches [each built to optimize the nature (density matrix or 
Hamiltonian) of the initial information] and give a deep insight in the physics of initial 
correlations. 
Indeed we can now express $\S_{\rm tot}$ in terms of $\S^{(r)}_{\rm 
tot}$ to 
obtain an {\em irreducible self-energy in terms of correlation 
blocks}. From
Eq. (\ref{eq5}) we have
\be
\S^{(r)}_{\rm tot}=\S_{\rm tot}\frac{1}{1-g\S_{\rm tot}}
\ee
and hence
\beq
\S_{\rm tot}&=&\S^{(r)}_{\rm tot}(1-g\S_{\rm tot})
\nonumber \\
&=&\S^{(r)}_{\rm tot}-\S^{(r)}_{\rm tot}g\S^{(r)}_{\rm tot}+\S^{(r)}_{\rm tot}
g\S^{(r)}_{\rm tot}g\S^{(r)}_{\rm tot}-\ldots\quad
\eeq
Inserting this expansion into 
Eq. (\ref{eq4}) we obtain a Dyson equation in which the irreducible 
self-energy depends only on $g$ and correlation blocks.

\section{Some applications and conclusions} 

The rules to expand the Green's function around any initial state constitute 
a powerful theoretical tool and the range of applicability can be vaster than we can imagine 
at present. Possible applications are in quantum chemistry where spin-degenerate ground-states 
are ubiquitous while the standard Wick theorem can only cope  with the highest spin component. 
The generalized Wick theorem is not limited to determinantal states thereby 
permitting to resolve the spin-multiplet properties.
This versatility can be used in the 
context of quantum entanglement as well. Another application is in the formalism of the rate equations 
which gives rise to negative time-dependent probabilities
when the standard Wick theorem is applied to  initially correlated states. Corrections with the 
$C_{l}$-blocks  should result in an 
improved theory. The generalized Wick theorem also allows us to address the effect of integrability-breaking 
perturbations using the exact Green's function of the integrable model.
More generally the generalized Wick theorem can be combined with any ground-state
numerical technique to calculate time-dependent properties.
Indeed from the ground-state density 
matrix we can calculate the non-equilibrium Green's function to 
any order in the interaction strength. 

The practical implementation of the generalized Wick theorem is no 
more difficult than the standard Wick theorem.
We expect that the generalized 
Wick theorem will be useful in a broad range of physical problems both in and 
out of equilibrium. We wish to conclude by observing that even though 
all derivations in this work pertain to Green's functions defined as 
averages over an initial state the generalization to Green's function 
defined as ensemble averages is straightforward and 
no extra complications arise.


\begin{thebibliography}{10}
\bibitem{deg-note}
We are implicitly assuming that the ground state is non-degenerate.
\bibitem{keldprop}
Note that $\S_{l}(1;x)$ and $\S_{r}(x;1)$ takes the same value for 
$z_{1}=t_{1\pm}$ on the two branches of the Keldysh contour.
\end{thebibliography}
\end{document}